\documentclass[pmlr,twocolumn,10pt]{jmlr} 

\usepackage{booktabs}
\usepackage{multirow}
\usepackage{array}
\newcolumntype{C}[1]{>{\centering\let\newline\\\arraybackslash\hspace{0pt}}m{#1}}

\usepackage{pifont}
\newcommand{\xmark}{\ding{55}}%

\usepackage{lipsum}

\usepackage{pgfplots}
    \pgfplotsset{compat=1.3}
\usepackage{adjustbox}

\usepackage[load-configurations=version-1]{siunitx} 

\theorembodyfont{\upshape}
\theoremheaderfont{\scshape}
\theorempostheader{:}
\theoremsep{\newline}

\jmlrvolume{LEAVE UNSET}
\jmlryear{2021}
\jmlrsubmitted{LEAVE UNSET}
\jmlrpublished{LEAVE UNSET}
\jmlrworkshop{Machine Learning for Health (ML4H) 2021} 

\title[Predicting Visual Improvement after Macular Hole Surgery]{Predicting Visual Improvement after Macular Hole Surgery: \titlebreak a Cautionary Tale on Deep Learning with Very Limited Data}

\author{%
\Name{Mathieu Godbout} \Email{mathieu.godbout.3@ulaval.ca}\\
\addr Université Laval
\AND
\Name{Alexandre Lachance} \Email{alexandre.lachance.7@ulaval.ca}\\
\addr Université Laval
\AND
\Name{Fares Antaki} \Email{fares.antaki@windowslive.com}\\
\addr Centre Hospitalier Universitaire de Montréal
\AND
\Name{Ali Dirani} \Email{drdirani@gmail.com}\\
\addr Université Laval
\AND
\Name{Audrey Durand} \Email{audrey.durand@ift.ulaval.ca}\\
\addr Université Laval
}

\begin{document}

\maketitle

\begin{abstract}
We investigate the potential of machine learning models for the prediction of visual improvement after macular hole surgery from preoperative data (retinal images and clinical features).
Collecting our own data for the task, we end up with only 121 total samples, putting our work in the very limited data regime.
We explore a variety of deep learning methods for limited data to train deep computer vision models, finding that all tested deep vision models are outperformed by a simple regression model on the clinical features.
We believe this is compelling evidence of the extreme difficulty of using deep learning on very limited data.
\end{abstract}

\begin{section}{Introduction}
\label{sec:intro}

Idiopathic full-thickness macular hole (MH) is a discontinuation of the neurosensory retina at the fovea and results in significant visual impairment, including reduced visual acuity (VA) and images distorsion.
Vitrectomy has been commonly used to treat MH, with recent studies reporting MH closure rate after a primary surgical procedure between 78 and 96\% \citep{yek2018outcomes,fallico2021factors,lachance2021revision}.  
Despite high closure rate, functional outcomes remain variable after successful surgery \citep{essex2018visual}.
The ability to predict whether a surgery would lead to a significant improvement on a patient's visual acuity before the actual surgery occurs would help clinicians suggest appropriate treatments and have a better understanding of prognostic factors.

Previous work attempted to predict the visual outcome after MH surgery using clinical factors (e.g. MH duration, preoperative VA, MH size) \citep{fallico2021factors,essex2018visual,steel2021factors} or more specific optical coherence tomography (OCT)-based features as macular hole index (MHI) \citep{geng2017area}.
Unfortunately, current predicting methods suffer from inaccuracy and variability.
Clinical factors are subjective and not always well standardized and OCT specific features do not account perfectly for the hole asymmetry and overall hole shape, reducing predictive potential \citep{murphy2020predicting}. 
OCT scans are cross-sectional images revealing the foveal and vitreous microstructure of a patient's eye and are commonly used in ophthalmology due to their noninvasive nature.
OCT scans are usually available in a 2D format and have been successfully used to detect various ocular diseases with deep computer vision models \citep{lu2018deep,li2019deep}.
This suggests that using raw OCT images may be sufficient for MH surgery outcome prediction.

To assess the potential of deep computer vision models on the task at hand, we build our own dataset.
Our dataset is composed of the preoperative information (OCT scans and clinical data) and postoperative visual outcome of 121 patients with successful MH surgery.
This amount of examples is drastically lower than what can be found in the popular ImageNet \citep{deng2009imagenet} dataset or other medical datasets \citep{khosravi2018robust,gulshan2016development,rajpurkar2017chexnet}, which range from a million to a few thousand images, respectively.
This very limited data regime is challenging for deep learning models.
In practice, transfer learning, where the model's weights are initialised from a presumably useful auxiliary task, is the standard way to circumvent this lack of training data.
However, recent work \citep{raghu2019transfusion} provided evidence that transfer learning is not necessarily useful on small medical datasets and that training smaller models from random initial weights is a promising alternative.

In this work, we investigate to which extent can deep learning help predict the visual outcome of MH surgery from OCT and clinical preoperative data.
Precisely, we wish to examine what training configurations lead to the best improvement on our very limited data regime.
To compare the usefulness of the deep computer vision models, we also train a baseline logistic regression on the tabular clinical data.
Lastly, we explore an approach combining information from OCT scans and clinical data to evaluate the potential information gain between both sources.
\end{section}

\begin{section}{Task description}
\label{sec:task_description}

We are interested in predicting the visual outcome of a patient from data available prior to a MH surgery.
While there are a few ways one can measure the visual outcome of the surgery, we choose to define it as an improvement of 15 letters on the ETDRS (Early Treatment Diabetic Retinopathy Study) visual acuity chart 6 months after the surgery. 
This VA gain of 15 ETDRS letters threshold is considered to represent a clinically significant improvement \citep{suner2009responsiveness}.
From a machine learning standpoint, our task is a binary classification problem, where a model is asked to predict a binary outcome, in our case whether the patient's VA will improve by at least 15 letters 6 months after surgery.

Our public dataset\footnote{Official link to be made available at the end of the anonymity period.} consists of 121 successful MH surgeries that occurred from 2014 to 2018 at the Centre Hospitalier Universitaire de Québec – Université Laval (CHU de Québec).
For each surgery, we possess two preoperative OCT scans of the operated eye, representing vertical and horizontal cuts.
The OCT scans are stored as high-definition \texttt{.tiff} files, each containing 750 x 500 pixels.
Alongside the OCT images, we also have access to clinical information of the patients, containing features such as MH duration and the patient's baseline preoperative VA.
All VA scores used were measured by the optometrists who did the operations for the study.
We randomly split our dataset in training, validation and test sets, each containing respectively 83, 21 and 17 patients.
More details on the dataset including a complete list of the available clinical features and some sample OCT scans can be found in \appendixref{apd:dataset_description}.

\end{section}

\begin{section}{Models}
\label{sec:models}

We divided all tested models in three categories: deep vision models (neural networks operating on OCTs only), logistic regression (operating on clinical data only) and a simple hybrid approach using both OCTs as well as clinical data.

\begin{subsection}{Deep Vision Models}

\textbf{ResNet-50.}
The first vision model we experiment with is the ResNet-50 \citep{he2016deep}.
It is very popular for transfer learning on medical images due to its successful applications on varied medical images like chest x-rays \citep{Wang_2017,farooq2020covid} or retinal fundus images \citep{shibata2018development}.
We distinguish two key hyperparameters when training ResNet-50 models: the model weight initialisation and the freezing of Convolutionnal Neural Network (CNN) weights.
For weight initialisation, we consider (1) a random initialisation, (2) an initialisation from the publicly available ImageNet pretrained weights and (3) weights obtained from the BYOL \citep{grill2020bootstrap} self-supervised weights pretraining on \citet{kermany2018identifying}'s dataset of around 85 000 OCTs of various eye pathologies. 
The freezing of CNN weights consists of only fine-tuning the linear prediction head on top of the model and is motivated when using pretrained CNN weights that can serve as good feature extractors on their own.

\noindent \textbf{CBR Family.}
The other vision models we use are the CBR family of models proposed by \citet{raghu2019transfusion}.
These 4 models (CBR-Tiny, CBR-small, CBR-Wide and CBR-Tall) contain only a fraction of the weights contained in standard ImageNet models and therefore represent minimal deep learning architectures.
Despite their small size, these models manage to reach or even surpass pretrained models like ResNet-50 on small medical image datasets \citep{raghu2019transfusion}, naturally making them interesting candidates for our dataset.

\end{subsection}

\begin{subsection}{Logistic Regression}
We implement a logistic regression baseline which is trained only on the clinical data to compare with the deep vision models trained on OCT images.
Our logistic regression model is implemented with $\ell_2$ regularization, where the regularization hyperparameter $C$ being automatically chosen from 5-fold cross validation.
The regression model is trained on the concatenation of the training and validation sets.
We applied a whitening transformation to the input to bring all features down to a normal distribution with 0 mean and unitary standard deviation.

\end{subsection}

\begin{subsection}{Combined approach}

In an attempt to see whether the combination of OCTs and clinical data can help boost prediction performance, we propose to use the late fusion \citep{huang2020fusion} technique to combine multimodal data.
Specifically, given a fully trained deep vision model, we extract the model's logistic prediction for a patient's OCT and concatenate it to the patient's clinical data.
This clinical data augmented with the vision model's predictions is then used to train another logistic regression, yielding an easy-to-implement way to combine OCTs and clinical data.
Moreover, the chosen late fusion technique also preserves model interpretability, a key concern in healthcare tasks.

\end{subsection}
\end{section}

\begin{section}{Experimental Setting}
\label{sec:exp_setting}

For our experiments with vision models, we leverage the fact that we have 2 OCTs for each patient to artificially \textit{double} our training samples by producing a training sample for each OCT.
At inference time, we simply take the average prediction from both OCTs associated with a patient.
We apply randomized data augmentation, randomly rotating, flipping horizontally and adjusting the brightness and contrast of every image in a training batch.
All augmented images are then resized to the 224 x 224 pixels range and are normalized using ImageNet's mean and standard deviation values.
Every vision model is trained using the binary cross-entropy loss with a batch size of 32, using the Adam \citep{kingma2014adam} optimizer with a learning rate of 0.0001 for a maximum of 1000 gradient steps.
We test the model's performance against the validation set every 50 steps and only retain the model with the highest validation AUROC.
Our codebase is publicly available\footnote{Official link coming soon.}.
Further implementation details can be found in \appendixref{apd:exp_settings}.

\end{section}

\begin{section}{Results}
\label{sec:results}

\begin{figure*}[htbp]
\floatconts
{fig:results_cnn}
{\caption{\textbf{Test set classification results for the vision models.}
We report the mean value and a 95 \% CI computed from $n=10$ independent runs.
ResNet-50 configurations are listed in the form \textit{RN-Init} for IN and BY the ImageNet and BYOL pretraining initializations, respectively.
CNN weights were frozen for models marked with $^\dagger$.
A table of these results is available in  \appendixref{apd:detailed_results}.}}
{
\begin{tikzpicture}
    \begin{axis}[
        height=0.35\textwidth,
        ybar=0.4pt,
        bar width=13pt,
        x=1.6cm,
        xtick distance=1,
        ytick={20,40,60,80},
        yticklabels={20\%, 40\%, 60\%, 80\%},
        xtick={1,2,3,4,5,6,7,8,9},
        xticklabels={RN-Rand\vphantom{RN-BY$^\dagger$},RN-IN\vphantom{RN-BY$^\dagger$},RN-IN$^\dagger$,RN-BY\vphantom{RN-BY$^\dagger$},RN-BY$^\dagger$,CBR-Tiny\vphantom{RN-BY$^\dagger$},CBR-Small\vphantom{RN-BY$^\dagger$},CBR-Wide\vphantom{RN-BY$^\dagger$},CBR-Tall\vphantom{RN-BY$^\dagger$}},
        xtick style={
            /pgfplots/major tick length=0pt,
        },
        x tick label style={font=\footnotesize},
        ytick pos=left,
        legend style={at={(0.25,0.98)},anchor=north west}
    ]
        \addplot+ [
        line width=0.4pt,
            error bars/.cd,
                y dir=both,
                y explicit,
                error bar style={line width=0.7pt},
                error mark options={
                rotate=90,
                      line width=0.7pt
                    },
        ] coordinates {
            (1,56.5) +- (0,22.7)
            (2,43.0) +- (0,14.7)
            (3,41.7) +- (0,13.5)
            (4,37.8) +- (0,11.4)
            (5,34.1) +- (0,22.0)
            (6,61.5) +- (0,23.7)
            (7,65.1) +- (0,21.8)
            (8,61.6) +- (0,11.1)
            (9,56.4) +- (0,28.6)
        };

        \addplot+ [
        line width=0.4pt,
            error bars/.cd,
                y dir=both,
                y explicit,
                 error bar style={line width=0.7pt},
                 error mark options={
                    rotate=90,
                      line width=0.7pt
                    },
        ] coordinates {
            (1,60.1) +- (0,18.1)
            (2,42.2) +- (0,10.9)
            (3,36.8) +- (0,13.7)
            (4,40.7) +- (0,18.0)
            (5,38.9) +- (0,10.8)
            (6,72.8) +- (0,14.6)
            (7,70.0) +- (0,18.5)
            (8,69.9) +- (0,20.4)
            (9,65.1) +- (0,23.3)
        };

        \legend{
            F1,
            AUROC,
        }
    \end{axis}
\end{tikzpicture}
}
\end{figure*}
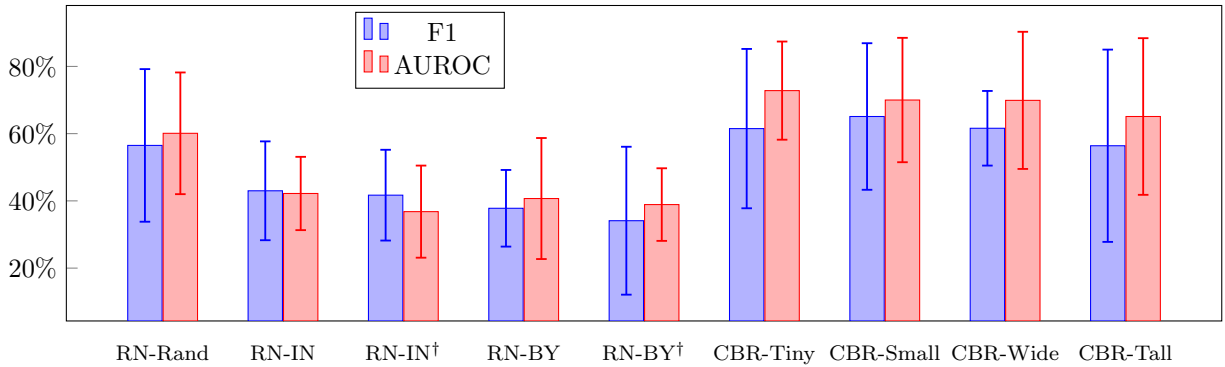

\noindent \textbf{Vision Models.}
\figureref{fig:results_cnn} presents the performance of the different vision models on the test set.
We highlight the overall poor performance (best average AUROC around 70\% and best F1 around 65\%) and high variance of all models (uncertainties in the 10-20 \% range).
We believe this can be attributed to the low data regime we are in giving too few training sample to enable adequate model generalization.
Nonetheless, we can still see that the CBR models consistently outperform the ResNet-50 models.
Surprisingly, the usage of both pretrained weights for the ResNet models only resulted in worse performance than random initialization, no matter if the weights were kept frozen or not.
Lastly, since all no single model outperforms the others in a statistically significant way, we opt to retain the CBR-Tiny model in following experiments because it is the one with the smallest number of parameters and should therefore be less prone to overfitting.

\noindent \textbf{Regression performance.}
\tableref{tab:reg_res} compares the performance of our best CNN model (CBR-Tiny) with the two logistic regression settings.
We see that the very simple logistic regression model performs surprisingly better than the CNN model, even though it is only fed clinical features.
Furthermore, using the predictions from OCTs only yields marginal, not statistically significant gains.
The slight increase in performance from using the OCT predictions might however indicate that there is relevant information in the OCT scans, but the trained CNN models are unable to extract it fully, likely due to the small amount of training examples.

\begin{table*}[hbpt]
\renewcommand{\arraystretch}{1.35}
\floatconts
{tab:reg_res}
{\caption{\textbf{Results comparison of the regression and CNN models.}
We report mean $\pm$ a 95 \% confidence interval for all results computed from $n=10$ independent runs.
Best means are \textbf{bolded}.}}
{
    \begin{tabular}{cccc}
        \toprule
         \textbf{Model}& \textbf{Input}& \textbf{F1} & \textbf{AUROC}\\
        \midrule
        Regression& Clinical data& 75.0 $\pm$ 0.0& \textbf{75.0} $\pm$ 0.0\\
        CNN& OCTs & 61.5 $\pm$ 23.7& 72.8 $\pm$ 14.6\\
        Regression + CNN& Clinical data + CNN predictions&\textbf{77.8} $\pm$ 4.3&\textbf{74.9} $\pm$ 7.0\\
        \bottomrule
    \end{tabular}
}
\end{table*}


\noindent \textbf{Model Interpretation.}
At last, we investigate what the regression model bases its prediction on.
To do so, we report the feature importance proportions in \figureref{fig:feat_importance} assigned by the regression model when trained with and without concatenating the CNN predictions.
Interestingly, the majority of the regression's weight is always put on the preoperative visual acuity.
Furthermore, we can also see that the CNN prediction is the feature with the least importance when introduced.
This last result seems to advise caution regarding the actual pertinence of using CNN predictions, as it indicates that CNN predictions do not bring much supplementary correlation with the target than the available clinical features.

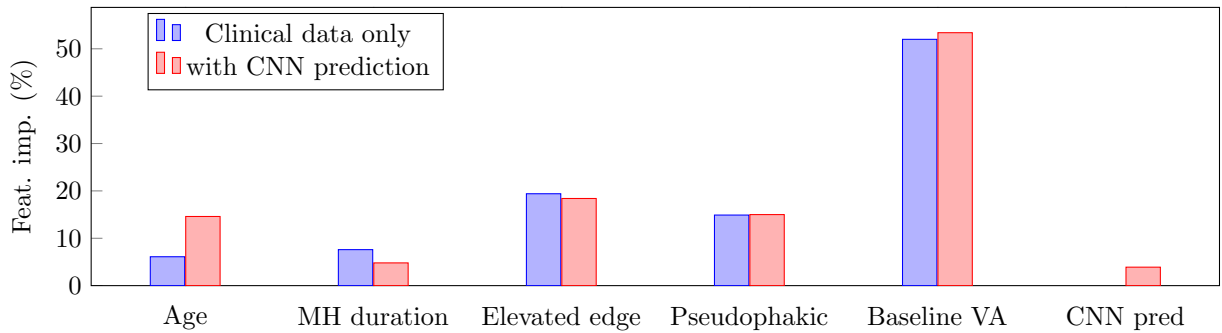
\begin{figure*}[hbtp]
\floatconts
{fig:feat_importance}
{\caption{\textbf{Mean feature importance assigned from a trained logistic regression.}
Here, MH stands for Macular Hole while VA stands for Visual Acuity.}}
{
\begin{tikzpicture}
    \begin{axis}[
        height=0.32\textwidth,
        ybar=0.4pt,
        ymin=0,
        bar width=13pt,
        x=2.5cm,
        xtick distance=1,
        ytick distance=10,
        ylabel=Feat. imp. (\%),
        xtick={1,2,3,4,5,6},
        xticklabels={Age, MH duration, Elevated edge, Pseudophakic, Baseline VA, CNN pred},
        xtick style={
            /pgfplots/major tick length=0pt,
        },
        ytick pos=left,
        legend style={at={(0.05,0.98)},anchor=north west}
    ]
        \addplot+ [
            line width=0.4 pt
        ] coordinates {
            (1,6.1) 
            (2,7.6) 
            (3,19.4) 
            (4,14.9) 
            (5,52.0) 
        };

        \addplot+ [
            line width=0.4 pt
        ] coordinates {
            (1,14.6) +- (0,18.1)
            (2,4.8) +- (0,10.9)
            (3,18.4) +- (0,13.7)
            (4,15.0) +- (0,18.0)
            (5,53.4) +- (0,10.8)
            (6,3.9) +- (0,14.6)
        };

        \legend{
            Clinical data only,
            with CNN prediction,
        }
    \end{axis}
\end{tikzpicture}
}
\end{figure*}



\end{section}

\begin{section}{Conclusion}

We built our own dataset of 121 patients with successful MH surgery to predict whether their visual acuity improved by at least 15 letters after 6 months.
Investigating if deep vision models trained on the patients' preoperative OCT scans can be useful, we find them to be marginally worse than a simple logistic trained on clinical data, despite resorting to several state-of-the-art tricks for dealing with limited data.
Moreover, we even find that logistic regression learns not to put much importance on predictions based on OCTs when given the opportunity.

\end{section}

\clearpage

\onecolumn

\bibliography{main}

\begin{thebibliography}{23}
\providecommand{\natexlab}[1]{#1}
\providecommand{\url}[1]{\texttt{#1}}
\expandafter\ifx\csname urlstyle\endcsname\relax
  \providecommand{\doi}[1]{doi: #1}\else
  \providecommand{\doi}{doi: \begingroup \urlstyle{rm}\Url}\fi

\bibitem[Deng et~al.(2009)Deng, Dong, Socher, Li, Li, and
  Fei-Fei]{deng2009imagenet}
Jia Deng, Wei Dong, Richard Socher, Li-Jia Li, Kai Li, and Li~Fei-Fei.
\newblock Imagenet: A large-scale hierarchical image database.
\newblock In \emph{2009 IEEE conference on computer vision and pattern
  recognition}, pages 248--255. Ieee, 2009.

\bibitem[Essex et~al.(2018)Essex, Hunyor, Moreno-Betancur, Yek, Kingston,
  Campbell, Connell, McAllister, Allen, Ambler, et~al.]{essex2018visual}
Rohan~W Essex, Alex~P Hunyor, Margarita Moreno-Betancur, John~TO Yek, Zabrina~S
  Kingston, William~G Campbell, Paul~P Connell, Ian~L McAllister, Penelope
  Allen, John Ambler, et~al.
\newblock The visual outcomes of macular hole surgery: a registry-based study
  by the australian and new zealand society of retinal specialists.
\newblock \emph{Ophthalmology Retina}, 2\penalty0 (11):\penalty0 1143--1151,
  2018.

\bibitem[Fallico et~al.(2021)Fallico, Jackson, Chronopoulos, Hattenbach, Longo,
  Bonfiglio, Russo, Avitabile, Parisi, Romano, et~al.]{fallico2021factors}
Matteo Fallico, Timothy~L Jackson, Argyrios Chronopoulos, Lars-Olof Hattenbach,
  Antonio Longo, Vincenza Bonfiglio, Andrea Russo, Teresio Avitabile, Francesca
  Parisi, Mario Romano, et~al.
\newblock Factors predicting normal visual acuity following anatomically
  successful macular hole surgery.
\newblock \emph{Acta Ophthalmologica}, 99\penalty0 (3):\penalty0 e324--e329,
  2021.

\bibitem[Farooq and Hafeez(2020)]{farooq2020covid}
Muhammad Farooq and Abdul Hafeez.
\newblock Covid-resnet: {A} deep learning framework for screening of {COVID19}
  from radiographs.
\newblock \emph{CoRR}, abs/2003.14395, 2020.
\newblock URL \url{https://arxiv.org/abs/2003.14395}.

\bibitem[Geng et~al.(2017)Geng, Wu, Jiang, Jiang, Zhu, Xu, Dong, and
  Yan]{geng2017area}
Xing-Yun Geng, Hui-Qun Wu, Jie-Hui Jiang, Kui Jiang, Jun Zhu, Yi~Xu, Jian-Cheng
  Dong, and Zhuang-Zhi Yan.
\newblock Area and volume ratios for prediction of visual outcome in idiopathic
  macular hole.
\newblock \emph{International journal of ophthalmology}, 10\penalty0
  (8):\penalty0 1255, 2017.

\bibitem[Grill et~al.(2020)Grill, Strub, Altch{\'{e}}, Tallec, Richemond,
  Buchatskaya, Doersch, Pires, Guo, Azar, Piot, Kavukcuoglu, Munos, and
  Valko]{grill2020bootstrap}
Jean{-}Bastien Grill, Florian Strub, Florent Altch{\'{e}}, Corentin Tallec,
  Pierre~H. Richemond, Elena Buchatskaya, Carl Doersch, Bernardo~{\'{A}}vila
  Pires, Zhaohan Guo, Mohammad~Gheshlaghi Azar, Bilal Piot, Koray Kavukcuoglu,
  R{\'{e}}mi Munos, and Michal Valko.
\newblock Bootstrap your own latent - {A} new approach to self-supervised
  learning.
\newblock In Hugo Larochelle, Marc'Aurelio Ranzato, Raia Hadsell,
  Maria{-}Florina Balcan, and Hsuan{-}Tien Lin, editors, \emph{Advances in
  Neural Information Processing Systems 33: Annual Conference on Neural
  Information Processing Systems 2020, NeurIPS 2020, December 6-12, 2020,
  virtual}, 2020.
\newblock URL
  \url{https://proceedings.neurips.cc/paper/2020/hash/f3ada80d5c4ee70142b17b8192b2958e-Abstract.html}.

\bibitem[Gulshan et~al.(2016)Gulshan, Peng, Coram, Stumpe, Wu, Narayanaswamy,
  Venugopalan, Widner, Madams, Cuadros, et~al.]{gulshan2016development}
Varun Gulshan, Lily Peng, Marc Coram, Martin~C Stumpe, Derek Wu, Arunachalam
  Narayanaswamy, Subhashini Venugopalan, Kasumi Widner, Tom Madams, Jorge
  Cuadros, et~al.
\newblock Development and validation of a deep learning algorithm for detection
  of diabetic retinopathy in retinal fundus photographs.
\newblock \emph{Jama}, 316\penalty0 (22):\penalty0 2402--2410, 2016.

\bibitem[He et~al.(2016)He, Zhang, Ren, and Sun]{he2016deep}
Kaiming He, Xiangyu Zhang, Shaoqing Ren, and Jian Sun.
\newblock Deep residual learning for image recognition.
\newblock In \emph{Proceedings of the IEEE conference on computer vision and
  pattern recognition}, pages 770--778, 2016.

\bibitem[Huang et~al.(2020)Huang, Pareek, Seyyedi, Banerjee, and
  Lungren]{huang2020fusion}
Shih-Cheng Huang, Anuj Pareek, Saeed Seyyedi, Imon Banerjee, and Matthew~P
  Lungren.
\newblock Fusion of medical imaging and electronic health records using deep
  learning: a systematic review and implementation guidelines.
\newblock \emph{NPJ digital medicine}, 3\penalty0 (1):\penalty0 1--9, 2020.

\bibitem[Kermany et~al.(2018)Kermany, Goldbaum, Cai, Valentim, Liang, Baxter,
  McKeown, Yang, Wu, Yan, et~al.]{kermany2018identifying}
Daniel~S Kermany, Michael Goldbaum, Wenjia Cai, Carolina~CS Valentim, Huiying
  Liang, Sally~L Baxter, Alex McKeown, Ge~Yang, Xiaokang Wu, Fangbing Yan,
  et~al.
\newblock Identifying medical diagnoses and treatable diseases by image-based
  deep learning.
\newblock \emph{Cell}, 172\penalty0 (5):\penalty0 1122--1131, 2018.

\bibitem[Khosravi et~al.(2018)Khosravi, Kazemi, Zhan, Toschi, Malmsten,
  Hickman, Meseguer, Rosenwaks, Elemento, Zaninovic,
  et~al.]{khosravi2018robust}
Pegah Khosravi, Ehsan Kazemi, Qiansheng Zhan, Marco Toschi, Jonas~E Malmsten,
  Cristina Hickman, Marcos Meseguer, Zev Rosenwaks, Olivier Elemento, Nikica
  Zaninovic, et~al.
\newblock Robust automated assessment of human blastocyst quality using deep
  learning.
\newblock \emph{bioRxiv}, page 394882, 2018.

\bibitem[Kingma and Ba(2015)]{kingma2014adam}
Diederik~P. Kingma and Jimmy Ba.
\newblock Adam: {A} method for stochastic optimization.
\newblock In Yoshua Bengio and Yann LeCun, editors, \emph{3rd International
  Conference on Learning Representations, {ICLR} 2015, San Diego, CA, USA, May
  7-9, 2015, Conference Track Proceedings}, 2015.
\newblock URL \url{http://arxiv.org/abs/1412.6980}.

\bibitem[Lachance et~al.(2021)Lachance, You, Garneau, Bourgault, Caissie,
  Tourville, and Dirani]{lachance2021revision}
Alexandre Lachance, Eunice You, J{\'e}r{\^o}me Garneau, Serge Bourgault,
  Mathieu Caissie, Eric Tourville, and Ali Dirani.
\newblock Revision surgery for idiopathic macular hole after failed primary
  vitrectomy.
\newblock \emph{Journal of Ophthalmology}, 2021, 2021.

\bibitem[Li et~al.(2019)Li, Chen, Liu, Zhang, Jiang, Wu, and Zhou]{li2019deep}
Feng Li, Hua Chen, Zheng Liu, Xue-dian Zhang, Min-shan Jiang, Zhi-zheng Wu, and
  Kai-qian Zhou.
\newblock Deep learning-based automated detection of retinal diseases using
  optical coherence tomography images.
\newblock \emph{Biomedical optics express}, 10\penalty0 (12):\penalty0
  6204--6226, 2019.

\bibitem[Lu et~al.(2018)Lu, Tong, Yu, Xing, Chen, and Shen]{lu2018deep}
Wei Lu, Yan Tong, Yue Yu, Yiqiao Xing, Changzheng Chen, and Yin Shen.
\newblock Deep learning-based automated classification of multi-categorical
  abnormalities from optical coherence tomography images.
\newblock \emph{Translational vision science \& technology}, 7\penalty0
  (6):\penalty0 41--41, 2018.

\bibitem[Murphy et~al.(2020)Murphy, Nasrulloh, Lendrem, Graziado, Alberti,
  La~Cour, Obara, and Steel]{murphy2020predicting}
Declan~C Murphy, Amar~V Nasrulloh, Clare Lendrem, Sara Graziado, Mark Alberti,
  Morten La~Cour, Boguslaw Obara, and David~HW Steel.
\newblock Predicting postoperative vision for macular hole with automated image
  analysis.
\newblock \emph{Ophthalmology Retina}, 4\penalty0 (12):\penalty0 1211--1213,
  2020.

\bibitem[Raghu et~al.(2019)Raghu, Zhang, Kleinberg, and
  Bengio]{raghu2019transfusion}
Maithra Raghu, Chiyuan Zhang, Jon Kleinberg, and Samy Bengio.
\newblock Transfusion: Understanding transfer learning for medical imaging.
\newblock In H.~Wallach, H.~Larochelle, A.~Beygelzimer, F.~d\textquotesingle
  Alch\'{e}-Buc, E.~Fox, and R.~Garnett, editors, \emph{Advances in Neural
  Information Processing Systems}, volume~32. Curran Associates, Inc., 2019.
\newblock URL
  \url{https://proceedings.neurips.cc/paper/2019/file/eb1e78328c46506b46a4ac4a1e378b91-Paper.pdf}.

\bibitem[Rajpurkar et~al.(2017)Rajpurkar, Irvin, Zhu, Yang, Mehta, Duan, Ding,
  Bagul, Langlotz, Shpanskaya, Lungren, and Ng]{rajpurkar2017chexnet}
Pranav Rajpurkar, Jeremy Irvin, Kaylie Zhu, Brandon Yang, Hershel Mehta, Tony
  Duan, Daisy~Yi Ding, Aarti Bagul, Curtis Langlotz, Katie~S. Shpanskaya,
  Matthew~P. Lungren, and Andrew~Y. Ng.
\newblock Chexnet: Radiologist-level pneumonia detection on chest x-rays with
  deep learning.
\newblock \emph{CoRR}, abs/1711.05225, 2017.
\newblock URL \url{http://arxiv.org/abs/1711.05225}.

\bibitem[Shibata et~al.(2018)Shibata, Tanito, Mitsuhashi, Fujino, Matsuura,
  Murata, and Asaoka]{shibata2018development}
Naoto Shibata, Masaki Tanito, Keita Mitsuhashi, Yuri Fujino, Masato Matsuura,
  Hiroshi Murata, and Ryo Asaoka.
\newblock Development of a deep residual learning algorithm to screen for
  glaucoma from fundus photography.
\newblock \emph{Scientific reports}, 8\penalty0 (1):\penalty0 1--9, 2018.

\bibitem[Steel et~al.(2021)Steel, Donachie, Aylward, Laidlaw, Williamson, and
  Yorston]{steel2021factors}
DH~Steel, PHJ Donachie, GW~Aylward, DA~Laidlaw, TH~Williamson, and D~Yorston.
\newblock Factors affecting anatomical and visual outcome after macular hole
  surgery: findings from a large prospective uk cohort.
\newblock \emph{Eye}, 35\penalty0 (1):\penalty0 316--325, 2021.

\bibitem[Suner et~al.(2009)Suner, Kokame, Yu, Ward, Dolan, and
  Bressler]{suner2009responsiveness}
Ivan~J Suner, Gregg~T Kokame, Elaine Yu, James Ward, Chantal Dolan, and Neil~M
  Bressler.
\newblock Responsiveness of nei vfq-25 to changes in visual acuity in
  neovascular amd: validation studies from two phase 3 clinical trials.
\newblock \emph{Investigative ophthalmology \& visual science}, 50\penalty0
  (8):\penalty0 3629--3635, 2009.

\bibitem[Wang et~al.(2017)Wang, Peng, Lu, Lu, Bagheri, and Summers]{Wang_2017}
Xiaosong Wang, Yifan Peng, Le~Lu, Zhiyong Lu, Mohammadhadi Bagheri, and
  Ronald~M. Summers.
\newblock Chestx-ray8: Hospital-scale chest x-ray database and benchmarks on
  weakly-supervised classification and localization of common thorax diseases.
\newblock \emph{2017 IEEE Conference on Computer Vision and Pattern Recognition
  (CVPR)}, Jul 2017.
\newblock \doi{10.1109/cvpr.2017.369}.
\newblock URL \url{http://dx.doi.org/10.1109/CVPR.2017.369}.

\bibitem[Yek et~al.(2018)Yek, Hunyor, Campbell, McAllister, Essex, Luckie,
  Hunyor, Chang, Kwan, Clark, et~al.]{yek2018outcomes}
John~TO Yek, Alex~P Hunyor, William~G Campbell, Ian~L McAllister, Rohan~W
  Essex, Alan Luckie, Alex Hunyor, Andrew Chang, Anthony Kwan, Ben Clark,
  et~al.
\newblock Outcomes of eyes with failed primary surgery for idiopathic macular
  hole.
\newblock \emph{Ophthalmology Retina}, 2\penalty0 (8):\penalty0 757--764, 2018.

\end{thebibliography}

\clearpage 

\appendix

\section{Dataset description}
\label{apd:dataset_description}

Our dataset consists of 121 successful\footnote{Success for the MH surgery represents effective eye closure. As explained in the introduction of the paper, macular hole closing does not necessarily lead to significant visual acuity improvement, hence our visual outcome prediciton task.} MH surgery that took place at the Cente Hospitalier Universitaire de Québec - Université Laval (CHU de Québec).
As is usual for machine learning needs, we split our dataset in training, validation and test subsets, respectively containing 83 (69~\%), 21 (17~\%) and 17 (14~\%) patients.

For each surgery, we have access to 2 preoperative high-definition OCT scans describing the macular hole alongside clinical information features.
\figureref{fig:oct_sample} displays two OCT scans samples, taken from patients with and without significant visual outcome 6 months after surgery.
\tableref{tab:dataset_distro} presents the mean and standard deviation of the different clinical features on each patients subset. 
The available clinical information features used are
\begin{itemize}
    \item \textit{Age}: Patient's age at the time of the surgery, in years.
    \item \textit{MH duration}: Duration in weeks between the first referral date and the surgery for a MH.
    \item \textit{Elevated edge}: Defined as the presence of elevated edges of neurosensory retina in relation to the retinal pigment epithelial plane.
    \item \textit{Pseudophakic}:  An eye with performed  cataract surgery; artificial lens implanted in an eye to replace the natural lens.
    \item \textit{Baseline VA}: Preoperative VA, measured in ETDRS letters.
\end{itemize}

\begin{figure}[htbp]
\floatconts
  {fig:oct_sample}
  {\caption{Example OCT scans taken from our dataset.}}
  {%
    \subfigure[An OCT scan where the patient's visual acuity \textbf{did} improve by 15 letters or more after 6 months.]{\label{fig:circle}%
      \includegraphics[width=0.45\linewidth]{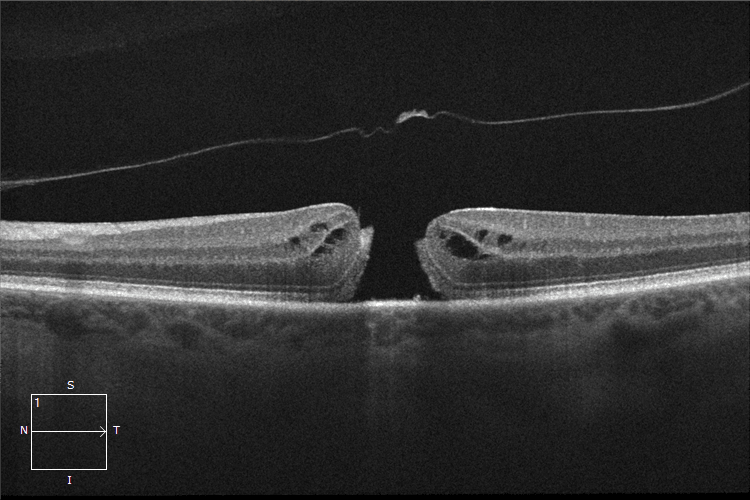}}%
    \qquad
    \subfigure[An OCT scan where the patient's visual acuity \textbf{did not} improve by 15 letters or more after 6 months.]{\label{fig:square}%
      \includegraphics[width=0.45\linewidth]{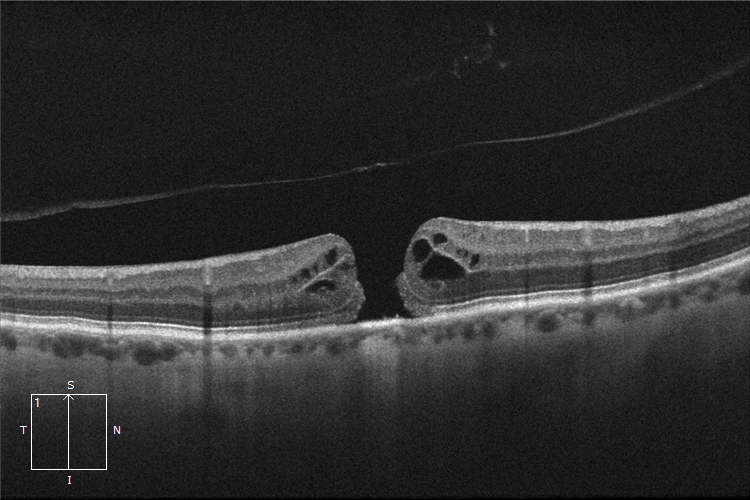}}
  }
\end{figure}

\begin{table*}[hbpt]
\renewcommand{\arraystretch}{1.25}
\floatconts
{tab:dataset_distro}
{
\caption{Baseline characteristics for patients of the different splits of the dataset. 
For real number characteristics, we report the mean and standard deviation values.
Here, MH stands for Macular Hole while VA stands for Visual Acuity.}
}
{
    \begin{tabular}{lC{3.5cm}C{3.5cm}C{3.5cm}}
        \toprule
         & \begin{tabular}{c}{}\textbf{Training set}\\($n=83$)\\\end{tabular}&\begin{tabular}{c}{}\textbf{Validation set}\\($n=21$)\\\end{tabular} & \begin{tabular}{c}{}\textbf{Test set}\\($n=17$)\\\end{tabular}\\
        \midrule
        Age: \textit{years} & 66.57 $\pm$ 7.60 & 65.19 $\pm$ 7.25 & 68.82 $\pm$ 6.72\\
        MH duration: \textit{weeks} & 11.48 $\pm$ 10.55& 9.95 $\pm$ 4.42& 10.31 $\pm$ 4.49\\
        Elevated edge: $n \,  (\%)$ & 74 (89.15)&18 (85.71)&13 (76.47)\\
        Pseudophakic: $n \,  (\%)$ & 14 (16.87)& 4 (19.05)& 4 (23.53)\\
        Baseline VA: \textit{letters} & 50.43 $\pm$ 15.51& 49.33 $\pm$ 13.53& 50.82 $\pm$ 17.62\\
        VA at 6 months: \textit{letters} & 66.01 $\pm$ 12.16& 65.10 $\pm$ 9.33& 66.00 $\pm$ 11.42\\
        VA gain $\geq 15$ letters: $n \,  (\%)$ & 41 (49.40)& 11 (52.38)& 8 (47.06)\\
        \bottomrule
    \end{tabular}
}
\end{table*}

\section{Detailed experimental setting}
\label{apd:exp_settings}

\textbf{Deep vision models.} 
All experiments were realised using a setup with a NVidia K80 GPU, an Intel Xeon Ivy Bridge E5-2697 v2 CPU and 12Gb of RAM.
All deep vision models training required less than 20 minutes in total.
The exact data augmentation for training is: (1) we resize images to the 256 x 256 pixels range, (2) we randomly flip the image on the horizontal with $p=0.5$, (3) we uniformly modify the image's brightness by a factor $\epsilon$ uniformly sampled in [-0.3, 0.3], (4) we uniformly modify the image's contrast by a factor $\epsilon$ uniformly sampled in [-0.3, 0.3], (5) we randomly rotate the image by $\theta$ degrees uniformly sampled in [-10, 10], (6) we take a random crop representing a ratio $p$ sampled from [0.7, 1.0] of the original image, (7) we resize the image to the 224 x 224 pixels range, (8) we normalize images using the ImageNet mean and standard deviations for all three RGB channels.
At inference time, we only resize the image to size 224 x 224 and apply ImageNet channel normalization.

\textbf{BYOL.} 
The BYOL self-supervised training is done with the publicly available implementation.
We scale back the implementation to our smaller scale, only training the model for 20 epochs and using a training batch size of 32, accumulating 8 batches before stepping, resulting in an effective batch size of 256.
We apply the same data augmentation as is used for our deep vision model training.

\textbf{Logistic regression.}
For logistic regression, we simply use the \texttt{LogisticRegressionCV} class from the high-quality scikit-learn library, keeping default values.
The whitening transformation is based on mean and standard deviation values of the regression set (containing both the training and validation subsets).

\section{Supplementary Results}
\label{apd:detailed_results}

We report in \tableref{tab:detailed_results} the complete results obtained for the deep vision models.
On top of the already reported F1 and AUROC metrics, we also present specificity and recall for each model.
\begin{table}[htbp]
\renewcommand{\arraystretch}{1.17}
\scriptsize
\floatconts
{tab:detailed_results}
{\caption{\textbf{Test set classification results for the vision models.}
We report the mean value and a 95 \% CI computed from $n=10$ independent runs.
$\dagger$ indicates whether or not the CNN weights were kept frozen during training.}}
{
    \begin{tabular}{ccccccc}
        \toprule
        \textbf{Model} & \textbf{Initialisation} & $\dagger$ & \textbf{F1} & \textbf{AUROC} & \textbf{Recall} & \textbf{Specificity}\\
        \midrule
        ResNet-50&	Random&	\xmark&	56.5 $\pm$ 22.7 (77.8)&	60.1 $\pm$ 18.1 (73.6)&	50.0 $\pm$ 28.3 (77.8)&	72.5 $\pm$ 21.6 (100.0)\\
        ResNet-50&	ImageNet&	\xmark&	43.0 $\pm$ 14.7 (55.6)&	42.2 $\pm$ 10.9 (50.0)&	42.2 $\pm$ 19.2 (55.6)&	41.2 $\pm$ 19.3 (62.5)\\
        ResNet-50 & ImageNet & \checkmark & 41.7 $\pm$ 13.5 (57.1)&	36.8 $\pm$ 13.7 (47.2)&	40.0 $\pm$ 20.2 (66.7)&	43.8 $\pm$ 27.7 (62.5)\\
        ResNet-50&	BYOL&	\xmark&	37.8 $\pm$ 11.4 (47.1)&	40.7 $\pm$ 18.0 (51.4)&	34.4 $\pm$ 11.8 (44.4)&	46.2 $\pm$ 22.3 (62.5)\\
        ResNet-50&	BYOL&	\checkmark&	34.1 $\pm$ 22.0 (55.6)&	38.9 $\pm$ 10.8 (47.2)&	31.1 $\pm$ 23.7 (55.6)&	46.2 $\pm$ 19.3 (62.5)\\
        CBR-Tiny&	Random&	\xmark&	61.5 $\pm$ 23.7 (77.8)&	\textbf{72.8} $\pm$ 14.6 (\textbf{87.5})&	57.8 $\pm$ 27.5 (77.8)&	67.5 $\pm$ 25.2 (87.5)\\
        CBR-Small&	Random&	\xmark&	\textbf{65.1} $\pm$ 21.8 (\textbf{88.9})&	70.0 $\pm$ 18.5 (80.6)&	\textbf{60.0} $\pm$ 22.4 (\textbf{88.9})&	\textbf{72.5} $\pm$ 34.7 (\textbf{100.0})\\
        CBR-LargeW&	Random&	\xmark&	61.6 $\pm$ 11.1 (66.7)&	69.9 $\pm$ 20.4 (84.7)&	56.7 $\pm$ 11.8 (66.7)&	68.8 $\pm$ 29.8 (87.5)\\
        CBR-LargeT&	Random&	\xmark&	56.4 $\pm$ 28.6 (77.8)&	65.1 $\pm$ 23.3 (79.2)&	53.3 $\pm$ 32.3 (77.8)&	62.5 $\pm$ 22.1 (75.0)\\
        \bottomrule
    \end{tabular}
}
\end{table}
\end{document}